\documentclass[12pt,thmsa]{article}
\usepackage{sw20lart}

\input tcilatex
\begin{document}

\author{Amjad Hussain Shah Gilani \\
National Center for Physics\\
Quaid-i-Azam University\\
Islamabad 45320, Pakistan\\
Email: ahgilani@yahoo.com}
\title{Why does set theory necessary to describe charge structure of particles ?}
\date{}
\maketitle

\begin{abstract}
It is pointed out that the set theory gave the exact symmetry while the
group theory did not. The triplicity of quarks and leptons is also pointed
out. The reason of seven families of particles and in each family eight
number of particles is elaborated.
\end{abstract}

\section{Introduction}

A group property violation in the charge structure of the gluons \cite
{hep-ph/0404026} changes \cite{hep-ph/0410207,hep-ph/0501103,hep-ph/0502055}
the entire scenario of the well established theory of strong interations 
\cite{pt2000aug} i.e. Quantum Choromodynamics (QCD). Many of the
experimental results \cite{hep-ph/9706416,PLB428-383,PLB384-388} are only
confirmed by the failure of perturbative QCD \cite{PLB384-388}. An
interacting gluon model is reviewed in Ref. \cite{hep-ph/0412293}, in which
the failure of perturbative QCD is elaborated. The exploration of physics
with $b$-flavoured hadrons offers a very fertile testing ground for the
standard model (SM) description of electroweak interactions \cite
{hep-ph/0003238,PRL10-531,PTP49-652,hep-ph/9803501,hep-ph/0304132}. The
theoretical \cite{hep-ph/0105302,hep-ph/0409133} and experimental \cite
{PRL84-5283,PRL89-231801,hep-ex/0308021,hep-ex/0408138} results for the
radiative $B$-decays to kaons resonances are quite opposite \cite
{hep-ph/0409133}. Some of the possible candidates of the discrepancy are
discussed in Ref. \cite{hep-ph/0409133}.

Issue of charge structure of fundamental particles has been discussed in the
recent articles \cite{hep-ph/0404026,hep-ph/0410207,hep-ph/0501103}. The
failure of group theory in predicting the charge structure of particles is
elaborated in Ref. \cite{hep-ph/0502055}. It is also pointed out that the
group theory can only provide constraints \cite{hep-ph/0502055}. But the
importance of set theory in prediction of charge structure of particles
discussed here.

\section{Where did we make a mistake ?}

Let us take nine unit cubes and make a symmetric pattern. The symmetric
pattern we observed, like $3^2$ (see Fig. 1 (a)). Take one cube away and we
have now eight cubes, like $3^2-1$ (see Fig. 1 (b)). We never make a
symmetric pattern on the surface with the help of eight cubes unless put
them upon each other and get a cubic pattern, like $2^3$ (see Fig. 1 (c)).

Quarks come in three colors, red $\left( r\right) $, blue $\left( b\right) $%
, and green $\left( g\right) $. QCD describes the interactions of colored
particles and such interactions are called chromodynamic interactions.
Chromodynamic interactions are mediated by gluons. Each gluon carries one
unit of color and one unit of anticolor. In terms of color SU$\left(
3\right) $ symmetry, we obtain nine states constitute a ``color octet'' and
a ``color singlet'' \cite{Griffiths1987}. ``color octet'' states are given
to gluons while the ``color singlet'' is thought to be as the photon. This
situation exactly resembled as shown in Fig. 1 (b). But we can never make an
exact symmetry of `8' on the surface which is explained by group theory. The
exact symmetry of `8' can only be taken in cubic form as predicted by set
theory \cite{hep-ph/0404026} and cube roots of unity \cite{hep-ph/0410207}
(see Fig. (c)).

\section{Another aspect}

The possible patterns with the help of eight cubes are shown in Fig. 2. The
Figs. 2(a--c) are the representations of $\left( 3^2-1\right) $ while Fig.
2(d) is the representation of $3^2-1=\left( 3-1\right) \left( 3+1\right)
=2\times 4=8$. The number of visual and hidden faces of the patterns of type 
$3^2-1$ and $2^3$ are listed in Table 1. If we look upon the Table 1, we see
that only Fig. 1(c) have equal number of visual and hidden faces of unit
cubes.

\begin{table}[tbp]
\caption{The number of visual and hidden faces of the patterns of type $3^2-1
$ and $2^3$.}\vspace{0.5cm}
\par
\begin{center}
\begin{tabular}{|c|c|c|}
\hline
& Visual faces & Hidden faces \\ \hline
Figure 2 (a) & 28 & 20 \\ \hline
Figure 2 (b) & 30 & 18 \\ \hline
Figure 2 (c) & 32 & 16 \\ \hline
Figure 2 (d) & 28 & 20 \\ \hline
Figure 1 (c) & 24 & 24 \\ \hline
\end{tabular}
\end{center}
\end{table}

\section{Why seven families of particles ?}

There are seven families of particles i.e., one gluon family \cite
{hep-ph/0410207}, three quark families and three lepton families \cite
{hep-ph/0501103}. Are there only seven families of particles or more ?. Yes,
there are only seven families of particles and no more. Why ? Let us have a
look on the cube. A cube has six faces and eight corners. So, a cube have
six face centers and one body center. The only one body center of the cube
resembles the gluon family while the six face centers resembles the
remaining six families of particles i.e. three quark families and three
lepton families.

The tiplicity of quarks and leptons has recently been discussed by Ma \cite
{hep-ph/0502024}. Ma proposed, how all three properties involving the number
three are connected in a fivefold application of the gauge symmetry SU$%
\left( 3\right) $. But we have different point of view. Let us look on Fig.
1(c), the $2^3$ has eight unit cubes and there 24 faces are visible and 24
faces are hidden. Three faces of each unit cube are visible and three are
hidden. We suspect that the visible faces gave the information about lepton
and their three families while the hidden faces gave the information about
quarks and their families.

\section{Conclusions}

We can never predict the volumetric properties of matter with the help of
group theory but only set theory can predict. We can only take the
constraints from the group theory. The set theory gave the exact symmetry
while the group theory did not. The triplicity of quarks and leptons is
discussed. The reason of seven families of particles and in each family
eight number of particles is elaborated.

These are the worries appeared from time to time during the review of Ref. 
\cite{hep-ph/0105302} discussed in Refs. \cite
{hep-ph/0404026,hep-ph/0410207,hep-ph/0501103,hep-ph/0502055} and in this
article.

\section{Figure Captions}

\begin{enumerate}
\item  The patterns of (a) $3^2$ (b) $3^2-1$ and (c) $2^3$.

\item  Possible diagrams for $3^2-1$.
\end{enumerate}

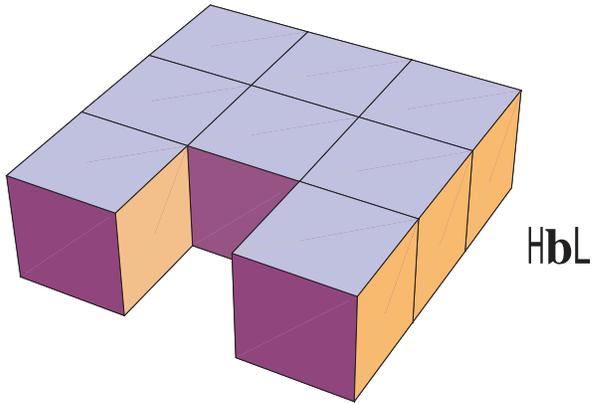
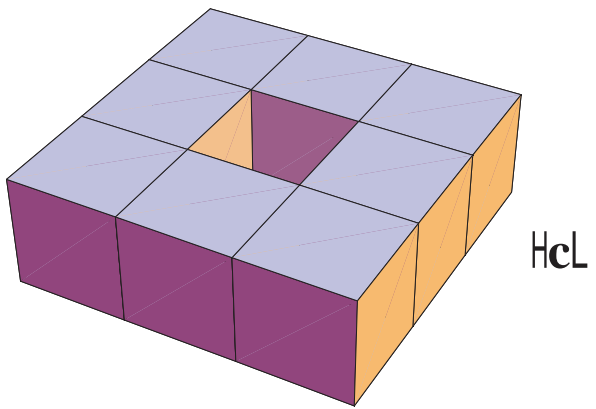
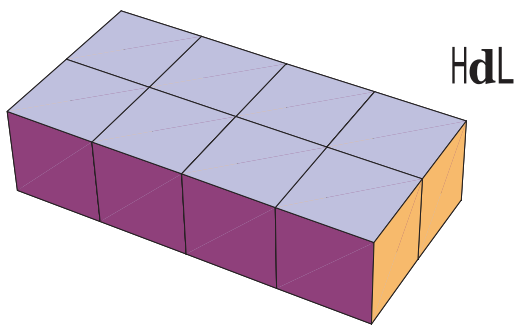
\end{document}